\documentstyle[prb,aps,epsf]{revtex}
\def\gsim{\,$\raise0.3ex\hbox{$>$}\llap{\lower0.8ex\hbox{$\sim$}}$\,}
\def\lsim{\,$\raise0.3ex\hbox{$<$}\llap{\lower0.8ex\hbox{$\sim$}}$\,}

\begin {document}

\draft

\twocolumn[\hsize\textwidth\columnwidth\hsize\csname@twocolumnfalse%
\endcsname

\title{Chiral order of Spin-1/2 frustrated quantum spin chains}
\author{T. Hikihara$^{1}$, M. Kaburagi$^{2,3}$ and H. Kawamura$^1$}
\address{$^1$Department of Earth and Space Science, 
Graduate School of Science, Osaka University, \\
Toyonaka, Osaka 560-0043, Japan\\
$^2$Faculty of Cross-Cultural Studies, 
Kobe University, Tsurukabuto, Nada, Kobe 657-8501, Japan\\
$^3$Graduate School of Science and Technology, Kobe University,
Rokkodai,  Kobe 657-8501, 
Japan}
\date{June 11, 2000}
\maketitle
\begin{abstract}
The ordering of
the frustrated $S$=$1/2$ $XY$ spin chain 
with the competing nearest- and next-nearest-neighbor 
antiferromagnetic
couplings, $J_1$ and $J_2$, is studied by using 
the density-matrix renormalization-group method. 
It is found that besides the
well-known spin-fluid and dimer phases the
chain exhibits  gapless `chiral' phase characterized
by the spontaneous breaking of parity, in which the
long-range order parameter is a chirality, 
$\kappa _l$=$S_l^xS_{l+1}^y-S_l^yS_{l+1}^x$, whereas 
the spin correlation decays algebraically. 
The dimer phase is realized for 
$0.33 \lsim j=J_2/J_1 \lsim 1.26$ while the chiral phase 
is realized for $j \gsim 1.26$.

\end{abstract}
]

Considerable attention has been paid to the ordering of 
frustrated quantum spin chains since these systems exhibit 
a wide variety of magnetic phases due to the interplay 
between quantum effect and frustration. 
We consider here an anisotropic frustrated quantum spin chain
described by the {\it XXZ\/} Hamiltonian,
\begin{equation}
   {\cal H} = \sum_{\rho=1}^{2}\Bigl\{J_{\rho}\sum_{\ell} 
       \bigl(S_\ell^x S_{\ell+\rho}^x + S_\ell^y S_{\ell+\rho}^y
           + \Delta S_\ell^z S_{\ell+\rho}^z\bigr)\Bigr\}~, 
\label{eq:Ham}
\end{equation}
where 
${\bf S}_{\ell}$ is the spin-$S$ 
spin operator at the $\ell$th site,
$J_{\rho} > 0$ is the antiferromagnetic interaction  
between the nearest-neighbor ($\rho$ = $1$) and the 
next-nearest-neighbor ($\rho$ = $2$) spin pairs, and 
$\Delta$ ($0 \le \Delta \le 1$) represents an exchange anisotropy.
Note that $\Delta\! = \!0$ and $\Delta\! = \!1$ represent the 
{\it XY\/} and Heisenberg chains, respectively.

In the classical limit, $S \to \infty$, the system in the ground state 
exhibits a magnetic long-range order (LRO) 
characterized by a wavenumber $q$.
The order parameter is defined by
\begin{equation}
\vec{m}(q) = \frac{1}{L} \sum_{\ell} \vec{S} e^{iq{\ell}}, \label{eq:hel}
\end{equation}
where $L$ is the total number of spins.
While the LRO is of the N{\`e}el-type ($q = \pi$)
when the frustration is smaller than a critical value, i.e.,
$j \equiv J_2/J_1 \le 1/4$, it becomes of helical-type  for $j > 1/4$
with
$q = \cos^{-1} \left( -1/4j \right) $.
It should be noticed that both the time-reversal and the parity symmetries are
broken in this helical ordered phase.
In the $XY$-like case ($0\leq \Delta < 1$), the helical ordered state possesses
a two-fold discrete chiral degeneracy characterized by the right- and
left-handed chirality, in addition to a continuous degeneracy
associated with the original $U(1)$ symmetry of the $XY$ spin.
The chiral order parameter is
defined by~\cite{chlOP}
\begin{eqnarray}
O_{\rm chiral} &=& \frac{1}{L} \sum_{\ell} \kappa_{\ell},\label{eq:chl} \\
\kappa_{\ell} &=& S_{\ell}^x S_{\ell+1}^y - S_{\ell}^y S_{\ell+1}^x
          = \left[ \vec{S}_{\ell} \times \vec{S}_{\ell+1} \right]_z.  \nonumber
\end{eqnarray}
The chiral order parameter changes its sign under the parity operation
but is invariant under the time-reversal operation.

In contrast to the classical case $S=\infty $,  
in the quantum case $S<\infty $
the magnetic LRO (\ref{eq:hel}) is completely
destroyed by strong quantum fluctuations.~\cite{Momoi}
Nevertheless, for 
the case of the $S=1$ $XY$-like chain, we have recently shown 
on the basis of
the exact-diagonalization (ED)
and the density-matrix renormalization-group (DMRG) 
methods that 
there exist novel chiral phases
in which only the chirality (\ref{eq:chl}) 
exhibits a LRO without the magnetic helical LRO (\ref{eq:hel}).~\cite{KKH} 
Two distinct chiral phases have been found, one gapless and the other gapped.
With increasing $j$, the system exhibits two phase transitions first
from the Haldane phase~\cite{Halconj} to the
gapped chiral phase, and then from the gapped chiral phase to the
gapless chiral phase. These chiral ordered phases
break only the parity symmetry spontaneously, with preserving both
the time-reversal and the translational symmetries. 
We also constructed the ground state phase 
diagram of the model (\ref{eq:Ham}) for the $S=1$ case.~\cite{HKKT,Hiki}

By contrast, the situation is less clear 
in the case of the $S=1/2$ $XY$ chain. 
There remains a discrepancy between 
the field-theoretical results~\cite{Ne} and  
our numerical results:~\cite{KKH} 
While the field-theoretical analysis  based on the
bosonization method predicted the occurrence of 
the chiral phase in the large $j$ region, 
the Binder parameter of the chirality calculated via the ED method 
up to $L=20$ exhibited no sign of the chiral phase.
Although Aligia {\it et al.\/}~\cite{ABE} recently pointed out 
that the system size $L=20$ might be insufficient 
to deal with the chirality in the large $j$ region,
the question whether 
the chiral phase is realized in the $S = 1/2$ $XY$ chain 
has not been clarified so far.

In the present paper, we try to
resolve this discrepancy by determining the phase diagram 
of the $S=1/2$ $XY$ chain (\ref{eq:Ham}) based on the
DMRG method.
The method used is the same as that in the previous work.~\cite{HKKT}
Using the DMRG method, we calculate
appropriate correlation functions associated with each
order parameter characterizing the phases, and
analyze their long-distance behaviors.
The results of our analysis strongly suggest 
the appearance of the chiral ordered phase for 
$j > j_c(\simeq 1.26)$.

Before going into details of our calculation, we briefly 
summarize the known properties of the ground-state phases of the 
$S = 1/2$ spin chain (1).
The system undergoes a phase transition
from the spin-fluid phase to the dimer phase at $j = j_d$
with increasing $j$.~\cite{Maj-Gho,Haldane,Tone-Hara,Oka-Nomu}
The critical value $j_d$ has been estimated to be
$j_d \simeq 0.33$ for the $XY$ case.~\cite{Oka-Nomu}
The spin-fluid phase realized at $j \le j_d$ 
is characterized by gapless
excitations above the ground state 
and an algebraic decay of spin correlations,
while the dimer phase realized at $j > j_d$ 
is characterized by a finite
energy gap above the doubly degenerate ground states and an
exponential decay of spin correlations. In the dimer phase, 
both the parity and the translational
symmetries are broken spontaneously.
The order parameter characterizing the dimer phase is given as
\begin{eqnarray}
O_{\rm dimer}^{\alpha} &=& \frac{1}{L} \sum_{\ell} \tau_{\ell}^{\alpha},
~~~~(\alpha = x,y,z)\label{eq:dim} \\
\tau_{\ell}^{\alpha} &=& (-1)^{\ell}S_{\ell}^{\alpha}S_{\ell+1}^{\alpha}. 
\nonumber
\end{eqnarray}
The dimer phase is further divided into two regions by the so-called
Lifshitz line at $j=j_L$ 
according to the nature of the spin correlations:
Whereas the structure factor $S(q)$ has a maximum 
at $q=\pi$ for $j \le j_L$, 
the wavenumber $q$ of the maximum of $S(q)$
shifts to an incommensurate value $q = Q < \pi$ for $j > j_L$.
This incommensurate character is regarded as the vestige of 
the helical order
in the classical case.~\cite{Roth}
Meanwhile, the question whether the chiral phase
exists for larger $j$ for the $XY$ chain still remains
controversial as mentioned above.

In order to probe the possible phases,
we calculate the  chiral, dimer and spin correlation functions
defined by
\begin{eqnarray}
C_\kappa (r) &=& \langle \kappa_{r_0-r/2} \kappa_{r_0+r/2} \rangle,
                   \label{eq:Cchl} \\
C_{\rm dim}^{x}(r) &=& \langle S_{r_0-r/2}^xS_{r_0-r/2+1}^x  
         \times (  S_{r_0+r/2}^xS_{r_0+r/2+1}^x \nonumber \\
&&                     - S_{r_0+r/2+1}^xS_{r_0+r/2+2}^x ) \rangle,
                   \label{eq:Cdim} \\
C_s^\alpha (r) &=& \langle S_{r_0-r/2}^\alpha S_{r_0+r/2}^\alpha \rangle
               ~~~(\alpha = x,z)  \label{eq:Ccor} 
\end{eqnarray}
which correspond to the order parameter (\ref{eq:chl}), 
(\ref{eq:dim}) and  (\ref{eq:hel}), respectively. 
Here $r_0$ represents the center position of open chain, i.e.,
$r_0 = L/2$ for even $r$ and $r_0 = (L+1)/2$ for odd $r$. We
employ the infinite-system DMRG algorithm
introduced by White.~\cite{White}
The notation $\langle \cdots \rangle $ represents the expectation value
in the lowest energy state in the subspace of $S_{total}^z = 0$.
Note that, since the chains treated in our calculation 
are sufficiently large,~\cite{Lconv} 
one can safely avoid the finite-size effect arising from
the incommensurate character of the spin correlation
as pointed out 
by Aligia {\it et al.}~\cite{ABE}

\begin{figure}
\begin{center}
\noindent
\epsfxsize=0.43\textwidth
\epsfbox{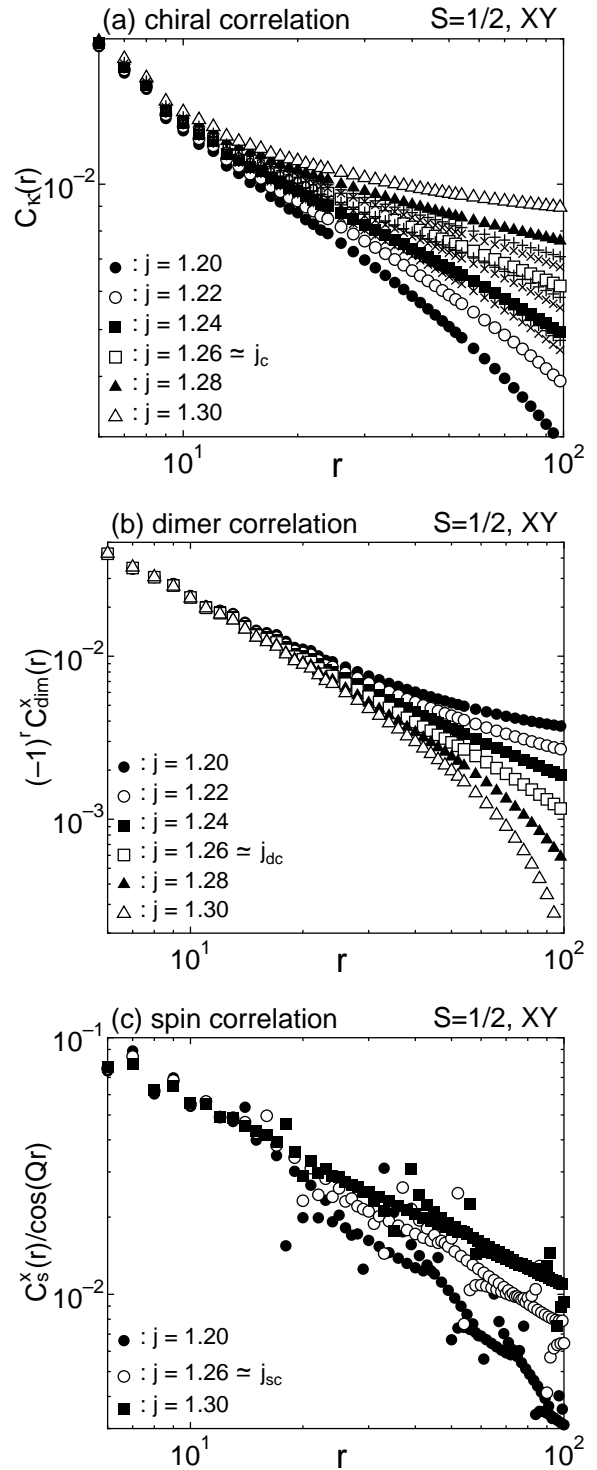}
\end{center}
\caption{Correlation functions versus $r$ on log-log plots
for various $j$: 
(a) chiral correlation $C_\kappa (r)$;
(b) dimer correlation $(-1)^r C_{\rm dim}^x(r)$;
(c) spin correlation $C_{s}^x(r)$ divided by the oscillating factor
$\cos (Qr)$. The number of block states kept in the DMRG
calculation is $m=450$.
To illustrate the $m$-dependence, we also
indicate by crosses the data for $m=400$ and $350$ for several cases 
where the $m$-dependence is relatively large.
In other cases, the truncation errors are smaller than the symbols.
}
\label{fig:DM}
\end{figure}

The calculated $r$-dependence of the chiral, dimer, and spin 
correlation functions 
are shown in Fig.\ref{fig:DM} (a)-(c) on log-log plots
for several typical values of $j$.
As can be seen from Fig.\ref{fig:DM} (a), the data of $C_\kappa (r)$
for $j > j_{c}\simeq 1.26$ are bent upward at larger $r$ suggesting a finite
chiral LRO, while for $j < j_{c}$ they are bent downward 
suggesting an 

\begin{figure}[ht]
\begin{center}
\noindent
\epsfxsize=0.43\textwidth
\epsfbox{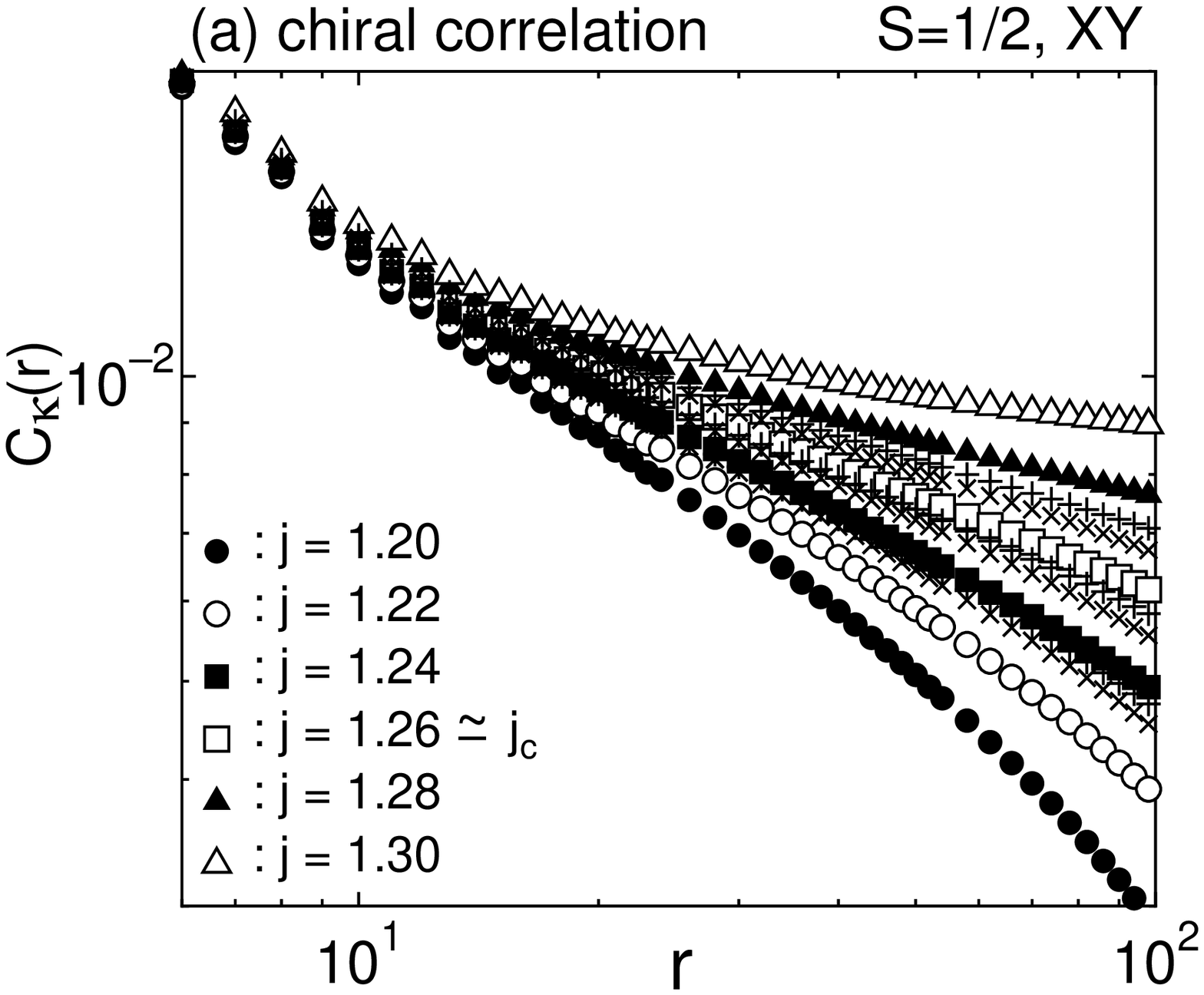}
\end{center}
\end{figure}

\vspace{-1.0cm}
\begin{figure}
\begin{center}
\noindent
\epsfxsize=0.43\textwidth
\epsfbox{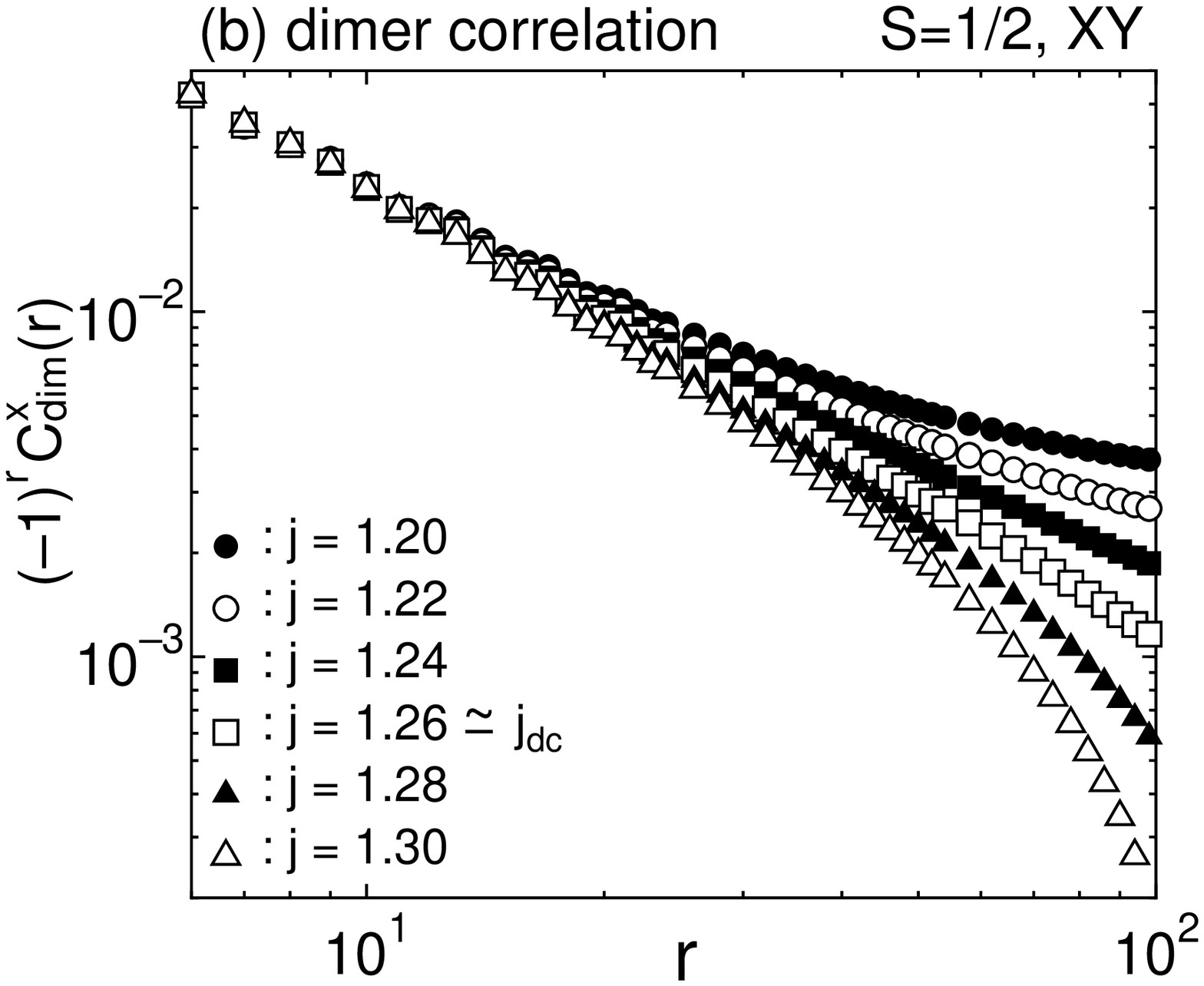}
\end{center}
\end{figure}

\vspace{-1.0cm}

\noindent
exponential decay of chiral correlations.
Although the data around the transition point suffer from 
the truncation error inherent to the DMRG calculation, 
we estimate the transition point as $j_c = 1.26^{+0.01}_{-0.03}$ 
by taking account of 
the $m$-dependence of the data shown in the figure.
The appearance of the chiral phase is consistent with 
the prediction of the field theory.~\cite{Ne,probED}

Meanwhile, as shown in Fig.\ref{fig:DM} (b), 
the data of $C_{\rm dim}^x (r)$ for $j < j_{dc} \simeq 1.26$ are 
bent upward for larger $r$ suggesting a finite
dimer LRO,
whereas they are bent downward for $j > j_{dc}$ suggesting 
an exponential decay of dimer correlations. 
The dimer transition point $j_{dc}$ is then estimated to be 
$j_{cd} = 1.26 \pm 0.01$. 
Figure \ref{fig:DM} (c) exhibits 
the spin correlation function $C_s^x(r)$ divided by 
the leading oscillating factor $\cos(Qr)$. 
The data of $C_s^x(r)$ are bent downward 
for $j < j_{sc} \simeq 1.26$ suggesting 
an exponential decay of spin correlations,
while  they exhibit a linear behavior for $j > j_{sc}$ 
suggesting a power-law decay of spin correlations.~\cite{scat}
This behavior of spin correlations
suggests that, as $j$ increases, the system exhibits a transition from 
the gapped state ($j < j_{sc}$) to the gapless state ($j > j_{sc}$).
We note that the absence of magnetic (spin) LRO has been proven 
rigorously for any $j$ and for general $S < \infty$.~\cite{Momoi}

The remaining problem is 
the relation among $j_c$, $j_{dc}$, and $j_{sc}$.
Two possibilities seem to be allowed from our data, {\it i.e.\/}, 
(i) $j_c = j_{dc} = j_{sc}$ 
or (ii) $j_c < j_{dc} = j_{sc}$.
If the case (i) is realized, the system undergoes only one phase
transition 
at $j = j_c = j_{dc} = j_{sc}$ between the dimer phase 
and the gapless chiral phase with no dimer order.
If the case (ii) is realized, on the other hand, 
the system undergoes two successive transitions on increasing $j$,
first at $j = j_c$ from the dimer phase to the ^^ ^^ chiral dimer" phase 
where both the dimer and chiral LRO's coexist with gapfull excitations,
and then at $j = j_{dc} = j_{sc}$ from 
the chiral dimer phase to the gapless chiral phase.
Although rather large error bars of $j_c$ and $j_{dc}$ 
prevent us from determining which of the cases is realized, 
our result suggests that the chiral dimer phase, 
if it ever exists, appears 
only in a rather narrow region, less than 3.2 \% of $j_c \simeq 1.26$, 
between the dimer and gapless chiral phases.
(We note that in the $S=1$ $XY$ chain the gapped chiral phase 
 exists for $0.473 \lsim j \lsim 0.49$,  whose width corresponds to 
 about 3.6 \% of $j_c \simeq 0.473$.~\cite{KKH,HKKT})
Further work will be necessary to solve 
the problem whether the chiral dimer phase exists.

In summary, from numerical studies of the ground-state properties of
the frustrated $S = 1/2$ $XY$ chain (\ref{eq:Ham}), 
we have found that 
\begin{enumerate}
\item{The chiral LRO is realized 
for a range of $j > j_c (\simeq 1.26)$.
This is consistent with the prediction 
by Nersesyan {\it et. al.\/}~\cite{Ne}}

\item{The dimer phase exists in the range $j_d (\simeq 0.33) 
< j < j_{dc} (\simeq 1.26)$.}
\end{enumerate}
Detailed results including  
the critical properties and the full phase diagram in the $j$-$\Delta$ plane 
will be reported elsewhere. 

We thank Prof. T. Tonegawa  for valuable discussion.
Numerical calculations were carried out in part at the Yukawa
Institute Computer Facility, Kyoto University.


\begin{thebibliography}{10}
\bibitem{Hiki}
The phase diagram in the $D$-$j$ plane for the $S=1$ case, where $D$ is 
the single-ion type anisotropy, is studied by A. Kolezhuk
(Phys. Rev. B {\bf 62}, R6057 (2000))
and by T. Hikihara (in this conference).
\bibitem{Lconv}
In the infinite-system DMRG algorithm, 
the system size increases by two at each DMRG step 
and the calculation is continued 
until the $L$-convergence of the data is attained.
We have performed the calculation more than 750 DMRG steps 
(corresponding to the system with $L=1500$ sites) 
and confirmed the $L$-convergence.
\bibitem{probED}
We consider that the failure of the ED analysis of Ref. \onlinecite{KKH} 
can be ascribed to the finite-size effect 
discussed in Ref. \onlinecite{ABE}.
The wavenumber $Q$ characterizing the incommensurability 
decreases rapidly from $Q = \pi$ at $j = j_L$ 
to $Q = \pi/2$ at $j \to \infty$ as $j$ increases.
In the $S=1/2$ $XY$ chain, the shift of $Q$ from $\pi/2$ 
turns out to be smaller than $0.03\pi$ for $j \gsim j_c \simeq 1.26$.
Meanwhile, in a finite open chain with $L$ sites, 
because of the condition that the wavefunction should vanish on both ends 
of the chain, 
the incommensurability smaller than $2\pi/L$ cannot be taken into account.
As a consequence, the system size must be larger than $L \sim 70$ 
if one wishes to detect the shift of $Q$ of order $0.03\pi$.
Thus, the calculation on chains up to $L=20$ might fail to 
extract the true asymptotic properties for $j \gsim j_c$.
\bibitem{scat}
The data of $C_s^x(r)$ shown in Fig.\ref{fig:DM} (c) 
are largely scattered, which 
might be attributed to the possible influence of correction terms 
characterized 
by  wavenumbers $Q' \ne Q$.
Nevertheless, we consider that we can judge the behavior of 
the leading term of $C_s^x(r)$ from the figure and 
make the conclusion in the text about $C_s^x(r)$.
\end{thebibliography}
\end{document}